\documentclass[reprint,amsmath, amssymb, aps, prd, superscriptaddress, nofootinbib, nobibnotes,longbibliography]{revtex4-1}

\usepackage{graphicx}
\graphicspath{{figs/}}
\usepackage{amsmath}
\usepackage{amssymb}
\usepackage{float}
\usepackage{xcolor}
\usepackage[colorlinks=true, allcolors=blue]{hyperref}
\usepackage{acro}
\usepackage{CJK}

\DeclareAcronym{GR}{
	short = GR,
	long  = general relativity
	}
\DeclareAcronym{BH}{
	short = BH,
	long  = black hole
}
\DeclareAcronym{BBH}{
	short = BBH,
	long  = binary black hole
}
\DeclareAcronym{BNS}{
	short = BNS,
	long  = binary neutron star
}
\DeclareAcronym{NSBH}{
	short = NSBH,
	long  = neutron-star-black-hole
}

\DeclareAcronym{GW}{
	short = GW,
	long  = gravitational wave
}

\DeclareAcronym{CBC}{
	short = CBC,
	long  = compact binary coalescence
}

\DeclareAcronym{PSD}{
	short = PSD,
	long  = power spectral density
}

\DeclareAcronym{ASD}{
	short = ASD,
	long  = amplitude spectral density
}

\DeclareAcronym{SNR}{
	short = SNR,
	long  = signal-to-noise ratio
}

\DeclareAcronym{ACF}{
	short = ACF,
	long  = auto correlation function
}

\DeclareAcronym{QNM}{
	short = QNM,
	long  = quasi-normal mode
}

\usepackage[capitalise]{cleveref}
\crefname{figure}{Fig.}{Figs.}
\Crefname{figure}{Fig.}{Figs.}

\def\be{\begin{equation}}
\def\ee{\end{equation}}
\def\({\left(}
\def\){\right)}
\def\[{\left[}
\def\]{\right]}
\def\msun{$M_\odot$}
\def\boldn{$\bold{n}$}

\def\msun{$M_\odot$}
\newcommand{\lmn}{{\ell m n}}

 \newcommand{\bqn}{\begin{eqnarray}}
 \newcommand{\eqn}{\end{eqnarray}}

\newcommand{\red}{\textcolor{black}}
\newcommand{\second}{\textcolor{black}}

\begin{document}
\begin{CJK*}{UTF8}{gbsn}
\title{A gating-and-inpainting perspective on GW150914 ringdown overtone: understanding the data analysis systematics}

\author{Yi-Fan Wang (王一帆)}
\email{yifan.wang@aei.mpg.de}
\affiliation{Max-Planck-Institut f{\"u}r Gravitationsphysik (Albert-Einstein-Institut), Am M{\"u}hlenberg 1, D-14476 Potsdam, Germany}
\affiliation{Max-Planck-Institut f{\"u}r Gravitationsphysik (Albert-Einstein-Institut), Callinstra{\ss}e 38, D-30167 Hannover, Germany}
\affiliation{Leibniz Universit{\"a}t Hannover, D-30167 Hannover, Germany}

\author{Collin D. Capano}
\affiliation{Department of Physics, Syracuse University, Syracuse, NY 13244, USA}
\affiliation{Department of Physics, University of Massachusetts, Dartmouth, MA 02747, USA}

\author{Jahed Abedi}
\affiliation{Beijing Institute of Mathematical Sciences and Applications (BIMSA), Huairou District, Beijing 101408, P. R. China}
\affiliation{Department of Mathematics and Physics, University of Stavanger, NO-4036 Stavanger, Norway}
\affiliation{Max-Planck-Institut f{\"u}r Gravitationsphysik (Albert-Einstein-Institut), Callinstra{\ss}e 38, D-30167 Hannover, Germany}
\affiliation{Leibniz Universit{\"a}t Hannover, D-30167 Hannover, Germany}

\author{Shilpa Kastha}
\affiliation{Saha Institute of Nuclear Physics, 1/AF Bidhannagar, Kolkata 700064, India}
\affiliation{Homi Bhabha National Institute, Training School Complex, Anushaktinagar, Mumbai 400094, India}
\affiliation{Niels Bohr International Academy, Niels Bohr Institute, Blegdamsvej 17, 2100 Copenhagen, Denmark}
\affiliation{Max-Planck-Institut f{\"u}r Gravitationsphysik (Albert-Einstein-Institut), Callinstra{\ss}e 38, D-30167 Hannover, Germany}
\affiliation{Leibniz Universit{\"a}t Hannover, D-30167 Hannover, Germany}

\author{Badri Krishnan}
\affiliation{Institute for Mathematics, Astrophysics and Particle Physics, Radboud University, Heyendaalseweg 135, 6525 AJ Nijmegen, The Netherlands}
\affiliation{Max-Planck-Institut f{\"u}r Gravitationsphysik (Albert-Einstein-Institut), Callinstra{\ss}e 38, D-30167 Hannover, Germany}
\affiliation{Leibniz Universit{\"a}t Hannover, D-30167 Hannover, Germany}

\author{Alex B. Nielsen}
\affiliation{Department of Mathematics and Physics, University of Stavanger, NO-4036 Stavanger, Norway}

\author{Alexander H. Nitz}
\affiliation{Department of Physics, Syracuse University, Syracuse, NY 13244, USA}

\author{Julian Westerweck}
\affiliation{Max-Planck-Institut f{\"u}r Gravitationsphysik (Albert-Einstein-Institut), Callinstra{\ss}e 38, D-30167 Hannover, Germany}
\affiliation{Leibniz Universit{\"a}t Hannover, D-30167 Hannover, Germany}
\affiliation{Institute for Gravitational Wave Astronomy and School of Physics and Astronomy, University of Birmingham, Edgbaston, Birmingham B15 2TT, United Kingdom}

\begin{abstract}
We revisit the recent debate on the evidence for an overtone in the black hole ringdown of GW150914 \red{using an independent data-analysis pipeline}. 
By gating and inpainting the data, we discard the contamination from earlier parts of the gravitational wave signal before ringdown. 
This enables parameter estimation to be conducted in the frequency domain, which is mathematically equivalent to the time domain method.
We keep the settings as similar as possible to the previous studies by~\textcite{Cotesta:2022pci} and Isi \textit{et al.}~\cite{Isi:2019aib,Isi:2022mhy} which yielded conflicting results on the Bayes factor of the overtone.
Our aim is to understand how different data analysis systematics, including sampling rates, erroneous timestamps, and the frequency resolution of the noise power spectrum, would influence the statistical significance of an overtone.
Our main results indicate the following: (i) a low-resolution estimation of the noise power spectrum tends to diminish the significance of overtones, (ii) adjusting the start time to a later digitized point reduces the significance of overtones, and (iii) overtone evidence varies with different sampling rates if the start time is too early, indicating that the overtone is a poor model, hence we propose a convergence test to verify the validity of an overtone model.
With these issues addressed, we find the Bayes factors for the overtone to range from $10$ to $26$ in a range of times centered at the best-fit merger time of GW150914, which supports the existence of an overtone in agreement with the conclusions of Isi \textit{et al.}~\cite{Isi:2019aib,Isi:2022mhy}.
\second{These results are obtained by keeping the start time and sky location fixed, enabling a direct comparison with other work. Marginalizing over these parameters would lower the Bayes factor to 1 for the evidence of an overtone.}
\end{abstract}

\maketitle
\end{CJK*}
\section{Introduction}
The \acp{GW} emitted during the \ac{BH} ringdown consists of a superposition of damped sinusoids known as \acp{QNM} \cite{qnm1}.
According to the no-hair theorem \cite{PhysRev.164.1776,PhysRevLett.26.331}, the characteristic frequencies and damping times are exclusively determined by the astrophysical \ac{BH}'s mass and spin.
When multiple modes are identified from the ringdown's \ac{GW}, the \ac{BH} mass and spin can be inferred independently and cross-checked. 
This is often known as \ac{BH} spectroscopy \cite{Dreyer:2003bv}, and offers an unequivocal way to test the validity of \ac{GR}. 

The first evidence of a \ac{QNM} is reported for GW150914 \cite{LIGOScientific:2016aoc,LIGOScientific:2016lio, LIGOScientific:2020tif}, in which the single $(\ell,m,n)=(2,2,0)$ mode is found with a frequency and decay time consistent with the \ac{GR} expectation from the full signal analysis.
Theoretical studies \cite{Cabero:2019zyt} suggest the detection of a secondary \ac{QNM} would likely only occur once Advanced LIGO \cite{TheLIGOScientific:2014jea}, Advanced Virgo \cite{VIRGO:2014yos} and KAGRA \cite{galaxies10030063} have reached their design sensitivities.
Nevertheless, the event GW190521 is discovered to have an unexpectedly high redshifted remnant mass $\sim$ 260 \msun{} \cite{Abbott:2020tfl,Abbott:2020mjq}.
The evidence of a subdominant mode $(3,3,0)$ from GW190521 is reported with a Bayes factor of 56 \cite{Capano:2021etf, Capano:2022zqm, Forteza:2022tgq, Abedi:2023kot}; also see \cite{Siegel:2023lxl} for an alternative interpretation involving the $(2,1,0)$ mode. 
By fitting numerical relativity data, Refs.~\cite{Giesler:2019uxc, Giesler:2024hcr} shows that the \ac{QNM} description can be valid as early as the merger stage, provided that overtones are considered, however, Ref.~\cite{Baibhav:2023clw} reports that the overtones identified in this region are not physical, \second{leaving it as an open question.}
Starting the analysis from the merger time, ~\textcite{Isi:2019aib} reports the first detection of a $(2,2,1)$ mode from GW150914 with a significance of $3.6 \sigma$ and shows the parameters to be consistent with the prediction of \ac{GR}.

However, \textcite{Cotesta:2022pci} claims that the detection of the overtone in ~\textcite{Isi:2019aib} is noise-dominated. 
\red{Their} reanalysis shows that the Bayes factor in favor of the overtone compared to only the fundamental mode is less than 1 around the merger time, hence no evidence for the overtone is found.
Nevertheless, \textcite{Isi:2022mhy} revisits the analysis and claims to be unable to reproduce the results in \textcite{Cotesta:2022pci} and shows that the Bayes factor of the model including the $(2,2,1)$ mode indicates the presence of the overtone.

The data analysis frameworks employed in \cite{Cotesta:2022pci} and \cite{Isi:2019aib,Isi:2022mhy} are fundamentally similar.
To remove the influence of the pre-ringdown \ac{GW} signal, both compute the likelihood in the time domain \cite{Carullo:2019flw, Isi:2021iql}, as \red{diagonalizing the covariance matrix with a Fourier transform} used by conventional \ac{GW} parameter estimation is no longer applicable.
In practice, \red{data analysis systematics, such as different data length and noise \ac{PSD} estimation} \cite{Isi:githubnote, reply_Carullo_etal}, can result in the inconsistent findings between \cite{Cotesta:2022pci} and \cite{Isi:2019aib,Isi:2022mhy}, highlighting the need for a comprehensive understanding of the \red{data analysis} techniques employed in ringdown overtone analysis.
In this study, we utilize an independent \red{gating-and-inpainting} framework to revisit GW150914, aiming to offer a new perspective to understand the analysis for overtones.
\red{Further research into the GW150914 overtone, such as \cite{CalderonBustillo:2020rmh, Finch:2022ynt, Ma:2023vvr, Ma:2023cwe, Crisostomi:2023tle}, utilizes various methodologies, however, we do not directly compare our findings with these studies since they essentially explore a different likelihood. 
We limit our comparative analysis to \cite{Cotesta:2022pci, Isi:2019aib, Isi:2022mhy,Wang:2023mst}, given that the underlying likelihoods are mathematically equivalent (see proof in \cref{sec:proof}). As a result, any observed discrepancies are attributable to data analysis systematics.}

In \cite{Capano:2021etf} a new approach is proposed to excise the contamination from the \ac{GW} signal prior to the ringdown by gating and inpainting the data following the original idea of \cite{Zackay:2019kkv}.
This is mathematically equivalent to the time domain method of \cite{Cotesta:2022pci,Isi:2019aib, Isi:2022mhy} and enables us to keep the analysis in the frequency domain.
Thereby, we can take advantage of the parameter estimation package \texttt{PyCBC inference} \cite{pycbcinference} by using several existing modules, such as data conditioning, \ac{PSD} estimation with Welch's method \cite{1967ITAE...15...70W, Allen:2005fk}, frequency-domain likelihood calculation \cite{speagle:2019}, which are well tested and were used in numerous previous studies, e.g., \cite{Nitz:2021uxj,Nitz:2021zwj}. 

We explore the impacts of different sampling rates on the GW150914 overtone significance as the sampling rate appears to differ between \cite{Cotesta:2022pci} and \cite{Isi:2019aib,Isi:2022mhy} (16384 Hz vs 2048 Hz). 
In addition, the extremely short decay time of the overtone, typically $\lesssim 1~\mathrm{ms}$, motivates us to examine the matched-filtering of high frequency contents of the signal, which is straightforward by examining the waveform in the frequency domain \red{(see \cref{sec:mf})}.
As emphasized by \cite{Cotesta:2022pci}, the merger time of GW150914 is subject to uncertainty, and we follow \cite{Cotesta:2022pci, Isi:2019aib, Isi:2022mhy} to select a set of discrete times centered around the best-fit merger time as the ringdown starting time.
\footnote{See \cite{Correia:2023bfn} which reports the analysis for GW150914 overtone by sampling and marginalizing over the sky location and starting time. 
\second{This marginalization would lower the Bayes factor to $1$.}}
We also carefully ensure the starting time is precisely implemented in \texttt{PyCBC inference}, rather than rounding up to that of a LIGO data sample, in light of our findings that the evidence of an overtone is highly sensitive to such approximations (see \cref{app:a}) because of the overtone's rapid decay. 
\red{We also find that the frequency resolution in the noise \ac{PSD} estimate affects the statistical significance of the (2,2,1) overtone (refer to \cref{sec:psd})}. 
We verify our method with numerical relativity simulations and report the results in \cref{sec:nr}.

\section{Gated Gaussian Likelihood}

We briefly review the gated Gaussian likelihood that employs data gating and inpainting \cite{Zackay:2019kkv,Capano:2021etf,Isi:2021iql}.
The conventional likelihood used in \ac{GW} parameter estimation for Gaussian and stationary noise is
\be
\label{eq:timedomain}
\mathcal{L}(\bold{n}) = \frac{1}{\sqrt{(2\pi)^N |C|}} \exp{\[-\frac{1}{2}\bold{n}^T C^{-1}\bold{n}\]}
\ee
where $\bold{n}$ is the noise vector with $N$ elements, $C$ is the covariance matrix of $\bold{n}$.
By the stationary assumption, $C$ is a Toeplitz matrix, and can be further diagonalized by a discrete Fourier transform basis matrix if the noise data is circulant.
Therefore, the likelihood can be greatly simplified in the frequency domain as
\be
\ln \mathcal{L}(\bold{n}) = -\frac{1}{2}\langle \bold{n}| \bold{n}  \rangle 
\red{ -\ln\sqrt{(2\pi)^N|C|}}
\ee
where the inner product is defined as
\be
\langle \bold{a} | \bold{b} \rangle = 
4 \Re \left\{
\frac{1}{T}\sum_{p=1}^{\lfloor \left( N-1 \right)/2 \rfloor} 
\frac{\tilde {\bold{a}}^\dagger_p \tilde{\bold{b}}_p}{S_{n,p}} 
\right\}
\ee
in which $\tilde{\bold{a}}$ is the discrete Fourier transform of $\bold{a}$ with $N$ samples across time $T$, and $\dagger$ means complex conjugation, $S_{n,p}(f)$ is the one-sided \ac{PSD} of the noise \second{and the subscript $p$ indicates the $p$-th elements.}
When a \ac{GW} is present, the noise can be obtained by subtracting the \ac{GW} waveform $\bold{h}$ from the detector measurement $\bold{d}$, so that the likelihood of a GW waveform is $\mathcal{L}(\bold{n}) = \mathcal{L}(\bold {d-h})$.

Ringdown inference aims to exclusively analyze signals after the remnant \ac{BH} enters a linear perturbation regime, hence the pre-ringdown contamination should be excised.
However, this would break the circularity condition due to the abrupt onset of the ringdown signal, thus the covariance matrix can not be diagonalized simply by a Fourier transform \cite{Isi:2021iql}. 
One needs to numerically invert the non-circulant covariance matrix in \cref{eq:timedomain}, as implemented by the time domain analysis \cite{Cotesta:2022pci,Isi:2019aib,Isi:2022mhy}.

Alternatively, Ref.~\cite{Zackay:2019kkv} proposes, and Ref.~\cite{Capano:2021etf} applies in ringdown analysis, a relation between the inversion of the covariance matrix from truncated data, $\bold{n}_\text{tr}$, and that from the complete data, $\bold{n}$, by replacing (inpainting) the excised data with $\bold x$.
Without loss of generality, we express the complete data $\bold n$ as the concatenation of three vectors, which is $\bold n = \bold n_1 \oplus \bold n_2 \oplus \bold n_3$, where $\oplus$ denotes the concatenation operation; the border of $\bold n_2$ and $\bold n_3$ delineates the pre-ringdown and ringdown stage, and $\bold n_2$ is long enough to cover the entire pre-ringdown \ac{GW} signals.
The truncated data can be expressed by $\bold n_\mathrm{tr} = \bold n_1 \oplus \bold n_3$.
The gated Gaussian likelihood aims to replace $\bold n_2$ with a vector $\bold x$, as $\bold n_\mathrm{inpaint} = \bold n_1 \oplus \bold x \oplus \bold n_3$, such that
\be
\label{eq:pretoeplitz}
\bold{n}_\mathrm{tr} C_\mathrm{tr}^{-1} \bold{n}_\mathrm{tr} =  {\bold n_\mathrm{inpaint}}^T C^{-1} \bold n_\mathrm{inpaint}
\ee
where $C_\mathrm{tr}^{-1}$ denotes the covariance matrix of $\bold n_\mathrm{tr}$.
The solution is obtained by solving the Toeplitz linear equation (for a proof see \cref{sec:proof})
\be
\label{eq:toeplitz}
\[ C^{-1}(\bold{n}_1 \oplus \bold{x} \oplus \bold n_3 ) \]_\mathrm{inpaint}= \bold{0}_\mathrm{inpaint}
\ee
where the subscript asserts this equation is valid only in the rows corresponding to the data being inpainted.
Given an $M$-dimensional inpainting vector $\bold x$, \cref{eq:toeplitz} is an $M$-dimensional Toeplitz linear equation with the time complexity scaling as $M^2$.
Since the right-hand side of \cref{eq:pretoeplitz} resumes the use of $C^{-1}$, one can diagonalize it with a discrete Fourier transform and thus perform the analysis in the frequency domain once $\bold x$ is obtained via \cref{eq:toeplitz}.

\section{Results of the overtone evidence}
\label{section:results}

We reanalyze the overtone of GW150914 data \cite{LIGOScientific:2019lzm} using similar settings to \cite{Cotesta:2022pci,Isi:2019aib,Isi:2022mhy}.
Our waveform model is
\be
h_+ + i h_\times = \sum_{\lmn} {}_{_{-2}}S_{\lmn} (\iota, \varphi;{\chi}_f) 
A_{\lmn} e^{i(\Omega_{\lmn}t +  \phi_{\lmn})} \, ,
\ee
where ${}_{_{-2}}S_{\lmn}$ are the spin-weighted spheroidal harmonics \cite{Chandrasekhar:1985kt, Berti:2005gp}, $\iota$ and $\varphi$ are the inclination angle and azimuthal angle; $\Omega_{\lmn} = 2\pi f_{\lmn} + i/\tau_{\lmn}$ is the complex frequency, $f_{\lmn}$ and $\tau_{\lmn}$ are the characteristic frequency and decay time exclusively determined by $M_f$ and $\chi_f$, the mass and spin of the remnant \ac{BH}; $A_{\lmn}$ and $\phi_{\lmn}$ are the amplitude and initial phase, which in principle can be determined by the initial conditions of the \ac{BH} perturbation, however, due to lack of concrete knowledge, we treat them as free parameters to be inferred from the data.

We follow Ref.~\cite{Cotesta:2022pci} and use the reference GPS time $t_\mathrm{ref} = 1126259462.42323$ s as a median of GW150914's merger time recorded by LIGO Hanford, and expand the analysis by scanning different starting times within $t_\mathrm{ref}\pm 1.5$ ms, corresponding to $2\sigma$ uncertainty around $t_\mathrm{ref}$.
\second{
We ensure that the ringdown analysis starts precisely at the designated time by reconstructing the sub-sampling data point by time-shifting in the frequency domain, instead of rounding it up to the nearest available discrete data point.
A comparison with not doing so can be found in \cref{app:a}.
}
The amplitude priors on $A_{220}$ and $A_{221}$ are uniform in $[0, 5\times 10^{-20}]$;
the phase priors on $\phi_{220}$ and $\phi_{221}$ are uniform in $[0,2\pi]$; all of which are identical to \cite{Cotesta:2022pci}.
The prior of final mass and final dimensionless spin is chosen to be uniform in [35,140] \msun{} and [0,0.99];
the inclination angle, azimuthal angle, polarization angle and sky localization are fixed to the values given by \cite{Cotesta:2022pci,Isi:2019aib,Isi:2022mhy}, which in turn are obtained from the maximum likelihood value from the analysis of the complete signal of GW150914.
\second{We use the \texttt{dynesty} sampler \cite{dynesty} to sample the likelihood  (note that its usage is not only restricted to a frequency domain likelihood).}
\second{The inspiral-merger-ringdown time length of the dominant mode of a 30-30 \msun{} binary is $ 0.94$ s with a lower frequency 20 Hz, hence we choose the gating-and-inpainting window to be $1~\text{s}$ to remove the pre-ringdown signal.}
We examine $8~\text{s}$ of data centered on the starting time, padded with an additional $4 \text{s}$ of data both at the beginning and end to address the boundary wraparound from the whitening filter, and taper the whitening filter to zero within the $4~\text{s}$ duration \cite{Allen:2005fk}. 
The autocorrelation function of the LIGO data generally decays to zero within a few seconds, as quantitatively demonstrated in Fig. 9 of \cite{Isi:2021iql}.
\second{The \cref{app:b} demonstrates the robustness of these choices by changing the gating length to $2~\text{s}$ or changing the data analysis duration from 8 s to $4~\text{s}$ (in addition to the padding time); we obtain consistent results.}
We also use four different sampling rates $f_s$ from 1024 Hz to 8192 Hz to study the impact of different upper frequency limits on overtone inference.
No higher sampling rate is considered because the LIGO data calibration is only valid from 10 Hz to 5 kHz \cite{LIGOScientific:2019hgc, LIGOScientific:2019lzm}.

\begin{figure}[htbp]
\includegraphics[width=\columnwidth]{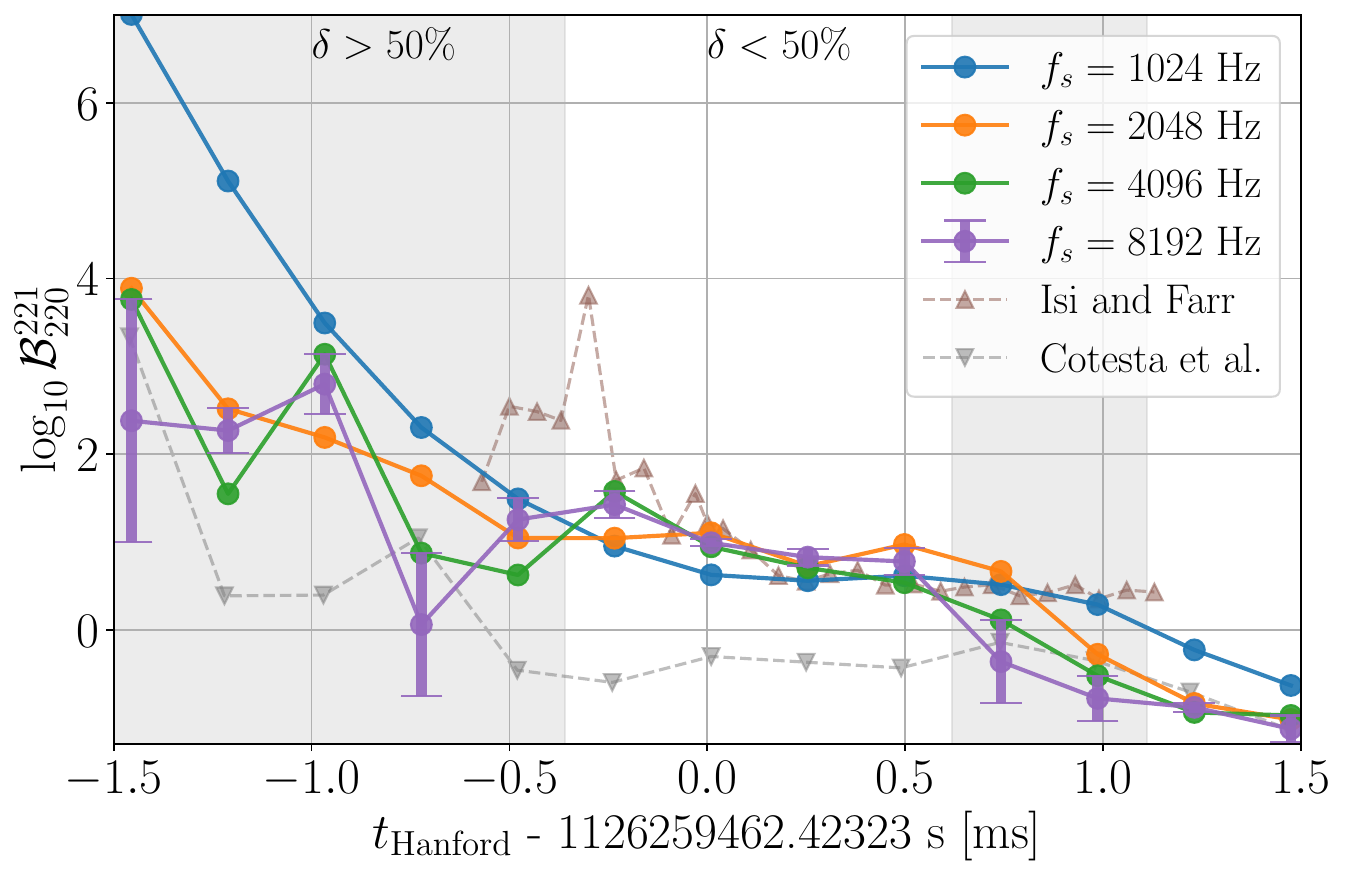}
\caption{The logarithm of Bayes factors comparing the $(2,2,0)+(2,2,1)$ model and the $(2,2,0)$-only model with respect to a variety of starting times for sampling rates $f_s=1024/2048/4096/8192$ Hz. As a comparison, we plot the Bayes factors obtained from \cite{Cotesta:2022pci,Isi:2022mhy}.
We also plot symmetric error bars with length $\mathcal{B}_{f_s = 8192}\times\delta$.
The shaded regions depict where the results from different sampling rates have not converged, quantified by $\delta > 50\%$, while in the non-shaded region all have $\delta < 50\%$.
}
\label{fig:bf}
\end{figure}

The findings of our study are presented in \cref{fig:bf}, which shows the logarithm of the Bayes factor $\log_{10} \mathcal{B}^{221}_{220}$, comparing the waveform model with modes $(2,2,0)+(2,2,1)$ to that with only the $(2,2,0)$ mode at various starting times.
\second{
We choose a stopping criterion for the \texttt{dynesty} sampler where the change in the natural logarithm of the Bayes factor is less than 0.1. Consequently, the uncertainty in the Bayes evidence ratio is approximately $2\log_{10}e^{0.1} \sim 0.08$ in \cref{fig:bf}, and therefore negligible.}
For comparison, we plot the Bayes factors reported by \cite{Cotesta:2022pci} and \cite{Isi:2022mhy}, respectively.
We consider the $f_s = 8192$ Hz runs as our fiducial results and quantify the convergence of Bayes factors by measuring the fractional difference between the highest and second highest sampling rate results, i.e., $\delta = |\mathcal{B}_{f_s = 8192}- \mathcal{B}_{f_s = 4096}|/\mathcal{B}_{f_s = 8192}$, as shown in \cref{fig:bf}.

We notice the intriguing trend that the Bayes factors from different sampling rates only start to converge after $t_\mathrm{ref}-0.25$ ms, where we quantify convergence by the criterion $\delta < 50\%$. 
Prior to that, there are noticeable disagreements from different sampling rates.
Results with $f_s=1024$ Hz yield the strongest evidence for the (2,2,1) mode (we will discuss in more details in the next section).
We regard the divergence as an indicator that the overtone model is matching the data insufficiently.
The strong Bayes evidence for low sampling rate can be plausibly attributed to matching the pre-ringdown stage of GW150914 which has a merger frequency $\sim 175$ Hz \cite{LIGOScientific:2016wyt,Carullo:2018gah}.
\second{As shown by \cref{fig:waveform}, the majority of the matched-filtering \ac{SNR}, defined as $\langle d | h \rangle / \sqrt{\langle h|h\rangle}$, would accumulate below $\sim$ 300 Hz for any sampling rates,} extending the high-frequency cutoff towards greater values would only result in increasing inconsistencies between the template and the signal.
In light of this observation, we propose a discriminator that utilizes the (non-)convergence of results from various sampling rates to determine the region where the ringdown overtone model is applicable, as opposed to the region where pre-ringdown contamination is present.
Signal consistency tests in a similar spirit have been proposed, e.g., for searching for \acp{GW} \cite{Allen:2004gu}.
At the late time around $t_\mathrm{ref}+1$ ms, we again observed discrepancies of different sampling rates, suggesting the overtone model is again not applicable.

Around the best-fit merger time of GW150914, specifically in $[-0.25,0.5]$ ms, we obtain converged Bayes factors from four different sampling rates consistently greater than 1 in favor of the existence of the $(2,2,1)$ mode.
In particular, at $t_\mathrm{ref} - 0.25$ ms, which was considered as the merger time by \textcite{Isi:2019aib}, we find $\mathcal{B}^{221}_{220}=26$, the median of $A_{221}$ deviates from zero with 2.8$\sigma$; at $t_\mathrm{ref}$, we find $\mathcal{B}^{221}_{220}=10$, and a 2.5$\sigma$ non-zero $A_{221}$, which indicates positive but moderate evidence for the presence of the $(2,2,1)$ mode.

Notably, our Bayes factors agree with those from \textcite{Isi:2022mhy} in and only in the convergence region $[-0.25,0.5]$ ms.
Nevertheless, there is a discrepancy at $-0.3$ ms with a notable outlier identified to have $\log_{10} \mathcal{B}\sim 4$ by \cite{Isi:2022mhy}.
Given their finer time stride, we further perform additional analyses with $f_s = 2048$ Hz with more finely spaced starting times, but can not reproduce the significant Bayes factor.
After $t_\mathrm{ref} + 0.75$ ms, our results are consistent with \textcite{Cotesta:2022pci}, indicating no overtone is found at a late time.
To understand how various methods affect the statistical significance of the overtone, we note that \red{one of the differences} is that \cite{Isi:2019aib, Isi:2022mhy} and \cite{Cotesta:2022pci} use a $0.2$ s and $0.1$ s duration for data analysis, respectively, while we use $16$ s to account for the non-zero \red{whitening filter} over a few seconds; thus we conclude a sufficiently long analysis duration can \red{be one of the factors to} enhance the statistical significance of finding an overtone.

\section{Understanding the discrepancy from different sampling rates}
\label{sec:mf}

\begin{figure}[htbp]
\includegraphics[width=\columnwidth]{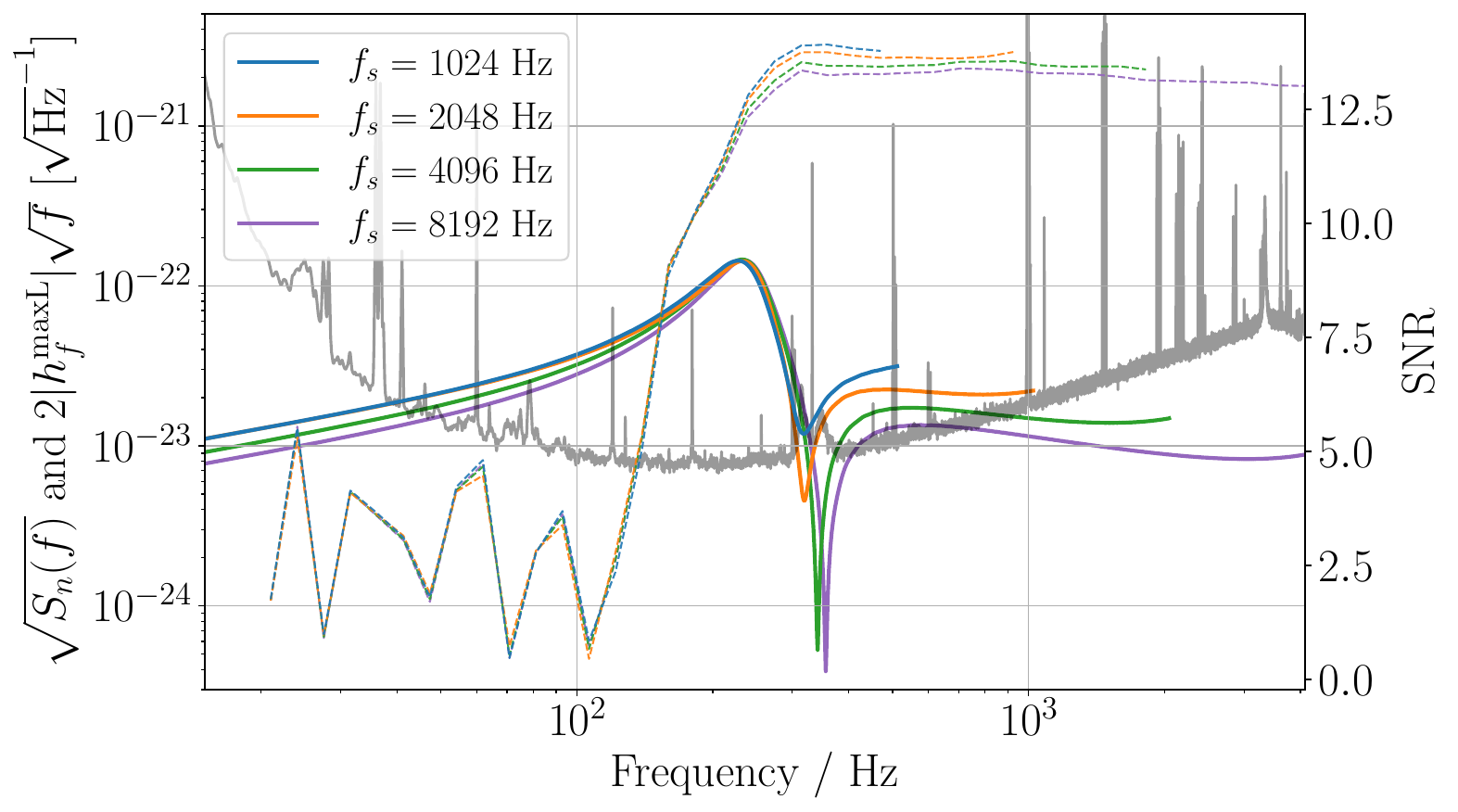}
\caption{The solid lines present the ringdown overtone waveform generated with the maximum likelihood parameters for four different sampling rates, with the scale shown on the left Y-axis.
The dashed lines present the matched-filtering \ac{SNR} as a function of the frequency upper limit of the integration.
The lower frequency limit is 20 Hz.
The ringdown starting time is $t_\mathrm{ref} - 0.75$ ms.
The grey curve shows the \ac{ASD} $\sqrt{S_n(f)}$.}
\label{fig:waveform}
\end{figure}

Prior to $t_\mathrm{ref} - 0.25$ ms, we notice a divergence of results from four different sampling rates.
To better understand its origin, we choose a particular time $t_\mathrm{ref} - 0.75$ ms, which shows discrepancies, and analyze the results of parameter estimation in depth.
In \cref{fig:waveform}, we present the Fourier transform of the ringdown overtone waveforms, $h_f^\mathrm{maxL}$, from parameters corresponding to the maximum likelihood sample.
We plot $2|h_f^\mathrm{maxL}|\sqrt{f}$ \second{(with scale shown on the left Y-axis)} such that the area under the square ratio between the waveform and the \ac{ASD} indicates the optimal \ac{SNR}, $\sqrt{\langle h|h\rangle}$.
\second{We also explicitly plot the matched-filtering \ac{SNR} accumulated from a lower frequency of 20 Hz up to a frequency upper limit shown in the X-axis, with the scale shown on the right Y-axis.
Also note that the waveforms shown in \cref{fig:waveform} are obtained by numerically Fourier transforming the ringdown overtone signals.
The Fourier transforms of the inspiral-merger-ringdown signals would fall off faster than any power law at high frequency, see e.g. Ref.~\cite{Husa:2015iqa}.
}

We notice that the $f_s = 1024$ Hz result tends to favor a waveform with higher amplitude at the Nyquist frequency $512$ Hz.
Examining the overtone amplitude $A_{221}$ reveals that a stronger $(2,2,1)$ mode is favored, which manifests as the tilt at high frequency due to the short decay time ($\sim1.5$ ms) of the overtone. 
The rapid decay leads to a broader frequency-domain representation of the waveform.
However, when the data analysis is extended to higher frequency bands, this strong (2,2,1) mode is no longer preferred \second{by the standard of Bayes factors.}
The \ac{SNR} also gradually decays from 13.76 to 13.71, 13.32, and 13.01 for sampling rates increasing from 1024 Hz to 8192 Hz.
The discrepancies suggest the starting time is too early and overtone templates do not match the data well.
The low sampling rate result tends to be more affected by the contamination from the pre-ringdown to produce a high \ac{SNR}.

\begin{figure}[htbp]
\begin{center}
\includegraphics[width=\columnwidth]{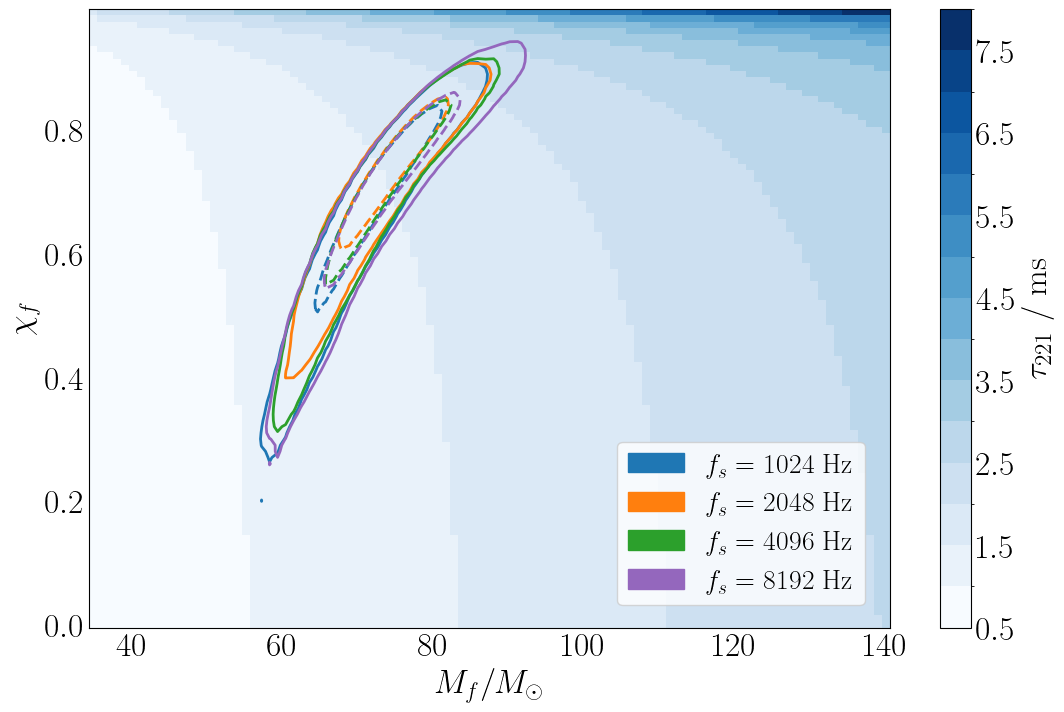}
\caption{The marginal posterior of $M_f$ and $\chi_f$ for four different sampling rates with analysis starting time $t_\mathrm{ref} - 0.75$ ms. 
In the background, we plot the value of $\tau_{221}$ as expected in \ac{GR} as a function of the mass and spin of a Kerr \ac{BH}.
The solid and dashed lines show the $90\%$ and $50\%$ credible regions, respectively.}
\label{fig:posmchi}
\end{center}
\end{figure}

\begin{figure}[htbp]
   \centering
\includegraphics[width=\columnwidth]{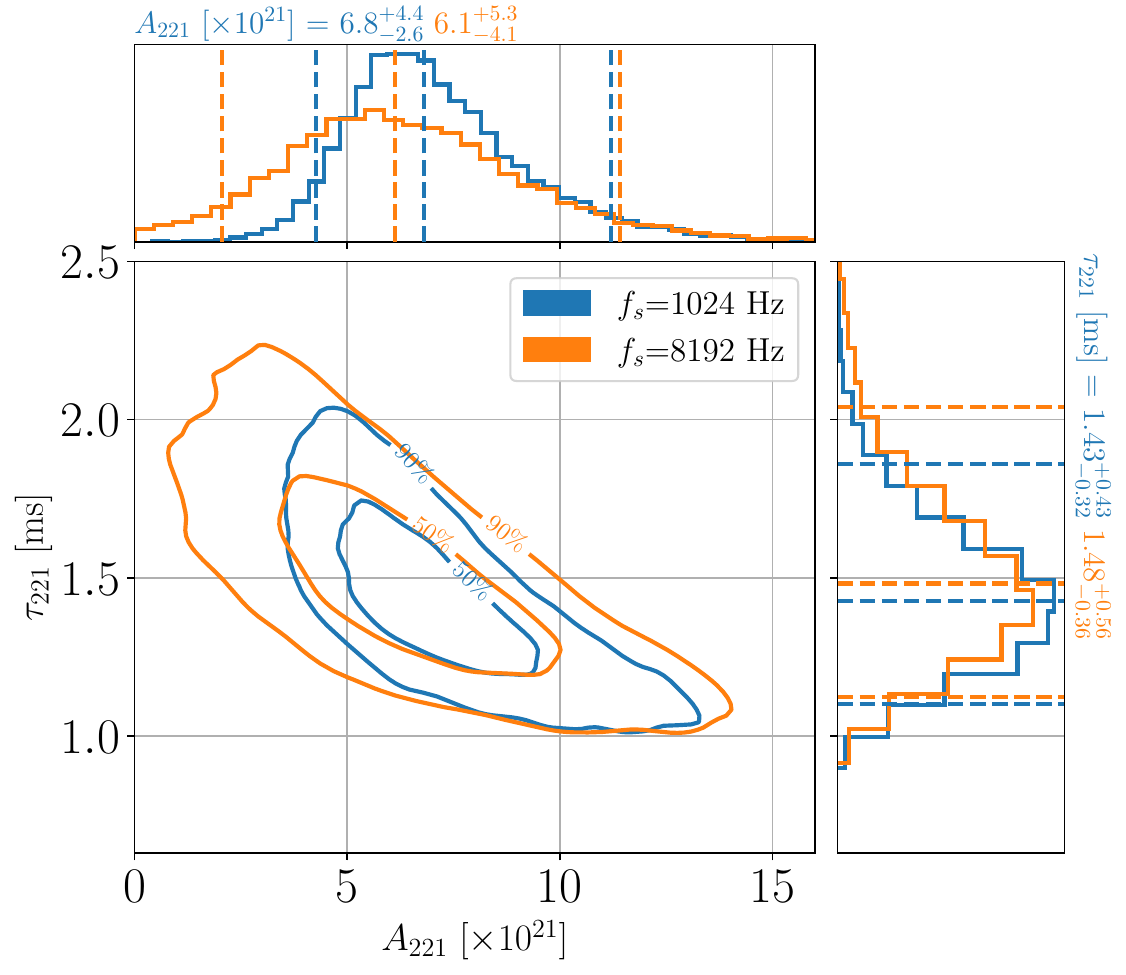}
\caption{A comparison of the posteriors for $A_{221}$ and $\tau_{221}$ using two different sampling rates, 1024 Hz and 8192 Hz, with analysis starting time $t_\mathrm{ref} - 0.75$ ms.
The contours denote the $90\%$ and $50\%$ credible regions. 
The shaded region shows the posterior probability density of the $f_s=1024$ Hz result.}
\label{fig:posatau}
\end{figure}

To further illustrate this with the entire posterior instead of a single point from maximum likelihood, we plot the mass and spin posterior distribution in \cref{fig:posmchi}.
For all $(M_f, \chi_f)$, we also compute the characteristic decay time of the overtone, $\tau_{221}$, predicted from \ac{GR} as the background of the figure.
\Cref{fig:posmchi} shows that the 8192 Hz result extends to a region with a longer decay time, while the low sampling rate results are more restricted to a shorter decay time.
The posterior of $\tau_{221}$ and $A_{221}$ is plotted in \cref{fig:posatau}, showing the parameters' negative correlation.
The 8192 Hz result favors a longer decay time $\tau_{221}$, hence a higher posterior density at $A_{221}=0$.
Using the Savage-Dickey density ratio, the density of $A_{221}=0$ directly determines $\mathcal{B}^{221}_{220}$.
We thus conclude that the (dis-) favoring of shorter $\tau_{221}$, which in turn is directly related to the signal intensity in high-frequency bands, by an inappropriate sampling rate can bias the posterior of $A_{221}$, and thus enhance or weaken the evidence for a $(2,2,1)$ mode.
The findings underscore the importance of examining the frequency spectrum in order to determine an appropriate time range within which a ringdown overtone model is applicable.

\section{Impact of an inaccurate starting time}
\label{app:a}

The original gating and inpainting formalism is potentially subject to a subtle caveat that the starting and ending time of the inpainting can only land on a specific data point due to the discrete nature of the sampled data.
As illustrated by \cref{fig:bf} and \cref{section:results}, the evidence of an overtone is sensitive to the ringdown starting time at a sub-millisecond level because of the rapid decay of an overtone.
This issue is particularly severe for a lower sampling rate with a coarser time resolution.
Consequently, it is necessary to ensure a precise starting time for the ringdown analysis.

\begin{figure}[htbp]
\includegraphics[width=\columnwidth]{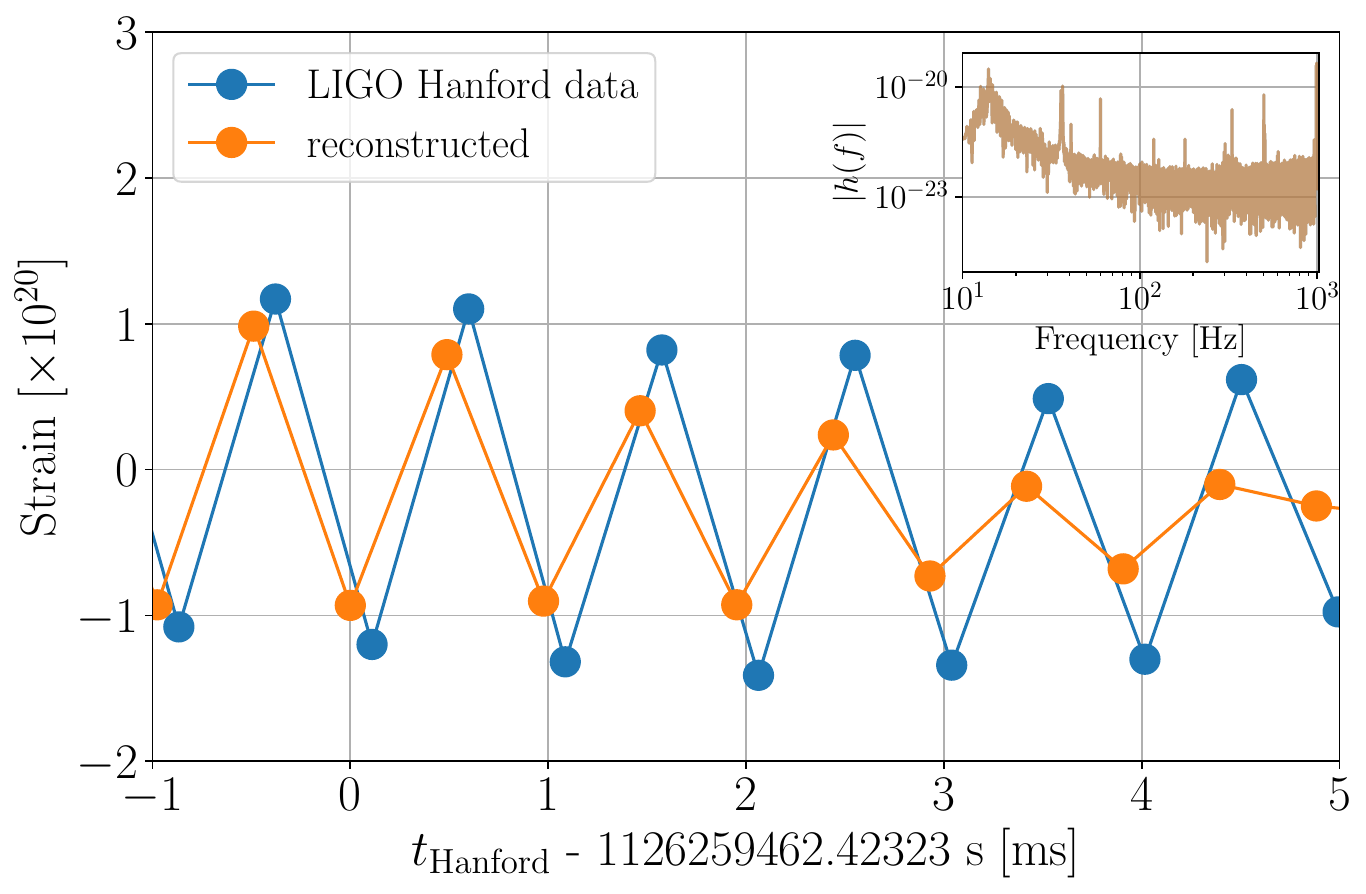}
\caption{Reconstructing the subsample corresponding to the starting time of ringdown, $t_\mathrm{ringdown}$. 
This is done by time shifting the data by an offset of the difference of $t_\mathrm{ringdown}$ and $t_\mathrm{nearest}$ which corresponds to the nearest data sample from LIGO Hanford.
\second{
In the inset we also plot the absolute value of the data before and after being time shifted in the frequency domain.
The figure shows they are identical as expected.}
}
\label{fig:timeshift}
\end{figure}

We have addressed this issue by reconstructing sub-data points from the sampled data at the starting time of ringdown, $t_\mathrm{ringdown}$, which is achieved by time shifting in the frequency domain by an offset between $t_\mathrm{ringdown}$ and the time stamp of the floor-nearest data sample, $t_\mathrm{nearest}$.
To visualize, the discrete data samples from the LIGO Hanford with $f_s = 2048$ Hz and those being reconstructed are plotted in \cref{fig:timeshift}.
\second{To verify this procedure, we plot the absolute value of the data before and after being time shifted in the frequency domain, they are expected to be identical and this is indeed the case as shown in the inset plot of \cref{fig:timeshift}.}

\begin{figure}[htbp]
\includegraphics[width=\columnwidth]{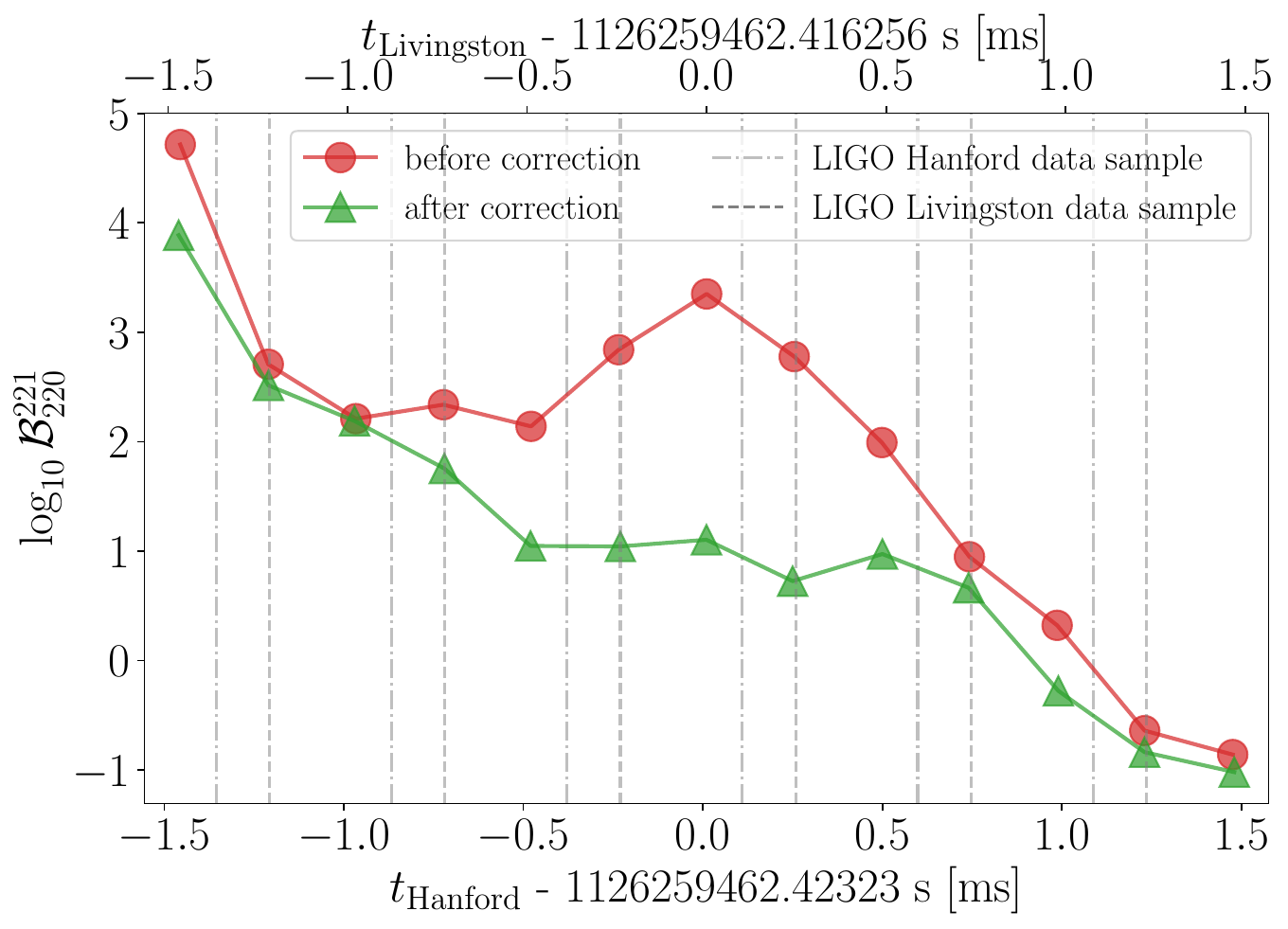}
\caption{The results of logarithm Bayes factor from sampling rate $2048$ Hz before and after accounting for the issue when the ringdown starting time lands on a subsample. 
The vertical (dot) dashed lines indicate the time stamp of the data samples from LIGO Hanford or Livingston.}
\label{fig:beforeafter}
\end{figure}

In \cref{fig:beforeafter} we plot the Bayes factor $\log_{10}\mathcal{B}^{221}_{220}$ from $f_s = 2048$ Hz before and after accounting for this issue.
To guide the eyes, we also plot vertical dashed lines for the time stamps for the sampled data of LIGO Hanford and Livingston.
Before addressing this problem, the starting time of ringdown, or equivalently, the ending time of inpainting, is rounded up to the floor-nearest time of a data sample.
This effectively results in an earlier and incoherent starting time between LIGO Hanford and Livingston, biasing the Bayes factors towards higher values due to the contamination from pre-ringdown signals.
Likewise, rounding up to a subsequent discretized data sample would reduce the overtone evidence. At a sampling rate of 16384 Hz, this effect should be insignificant (refer to Refs.~\cite{Isi:githubnote, reply_Carullo_etal} for further discussions on time discretization in time-domain data analysis pipelines). Nonetheless, our \cref{fig:beforeafter} illustrates that Bayes factors can be significantly biased by several orders of magnitude if an accurate starting time is not considered when using a 2048 Hz sampling rate.

\section{Impact of noise power spectrum estimation}
\label{sec:psd}

This section investigates the impact of different \ac{PSD} estimation techniques within our pipeline. 
According to Ref.~\cite{reply_Carullo_etal}, the choice of \ac{PSD} estimation method affects the overtone evidence utilizing either the frequency-domain Welch method \cite{1967ITAE...15...70W} or a time-domain \ac{ACF} based approach. 
However, we identified an additional systematic error: a lower frequency resolution in the noise \ac{PSD} estimation significantly diminishes the evidence for the (2,2,1) mode.

\begin{figure}[htbp]
\includegraphics[width=\columnwidth]{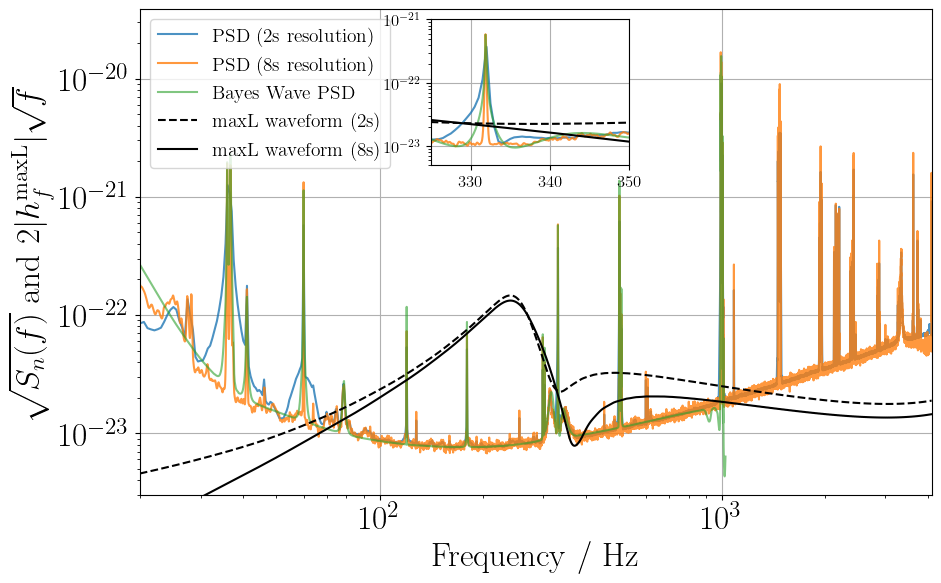}
\caption{Comparison of the different \ac{PSD} estimations obtained by this work using a resolution of $8 \text{s}$, that from \cite{Cotesta:2022pci} with a $2 \text{s}$ resolution, and the \ac{PSD} obtained by \texttt{BayesWave} used for GW150914 parameter estimation.
The maximum likelihood waveforms of an overtone model corresponding to the two \acp{PSD} are also plotted.}
\label{fig:psd}
\end{figure}

\Cref{fig:psd} shows the \ac{PSD} used in this study, estimated using the Welch method, which involves Fourier transforming several $8$-second data segments, totaling $512$ seconds, as described in Ref.~\cite{Allen:2005fk}.
For comparison, the \ac{PSD} from Ref.~\cite{Cotesta:2022pci} is plotted, which was derived using the authors' released configuration \cite{cotesta_release}, with $2$-second data segments in the Fourier transform.
Additionally, we present the \ac{PSD} estimated by the \texttt{BayesWave} algorithm \cite{2015CQGra..32m5012C, PhysRevD.91.084034, Cornish:2020dwh}, employed by the LIGO and Virgo collaboration for the GW150914 parameter estimation and publicly available \cite{bayeswavepsd}.
A notable feature is the enhanced resolution of a noise line structure in the \ac{PSD} used in this study, which we primarily attribute to the use of $8$-second data segments among other technical factors such as window functions.
On the other hand, we observed no visual differences between the \ac{PSD} obtained via the Welch method and that from numerically Fourier transforming a time-domain \ac{ACF}, so we plot only the result from the Welch method.

\begin{figure}[htbp]
\includegraphics[width=\columnwidth]{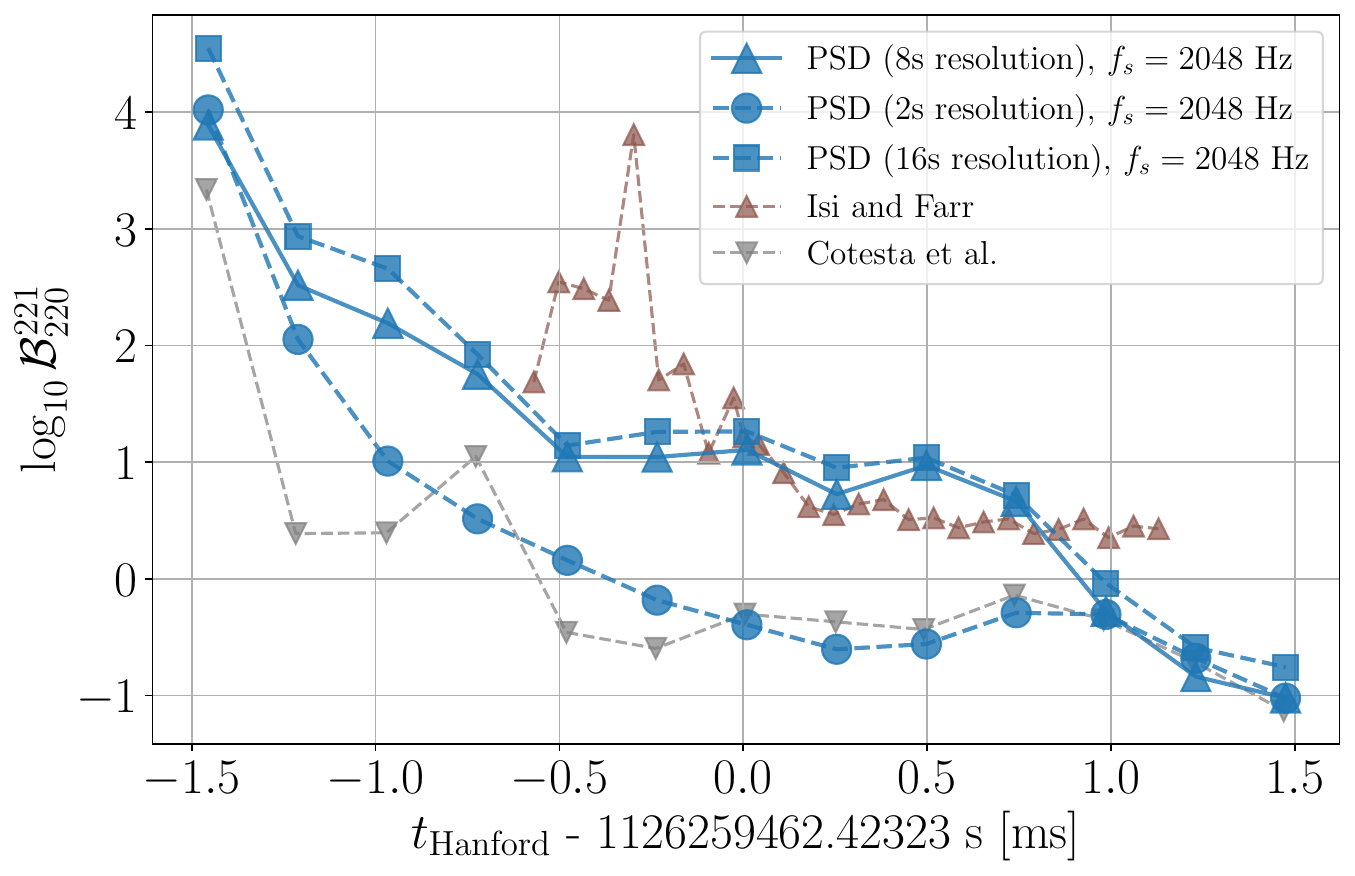}
\caption{The Bayes factors of a overtone model by using the \ac{PSD} of $16 \text{s}$, $8 \text{s}$ and $2 \text{s}$ resolution, respectively.
Here only shows the results from a sampling rate of 2048 Hz, and the results from other sampling rates are consistent.
The Bayes factors from \cite{Cotesta:2022pci,Isi:2022mhy} are plotted for comparison.}
\label{fig:psdbf}
\end{figure}

We conduct Bayesian parameter estimation for models both with and without an overtone, replacing the \ac{PSD} with an estimate from a $2 \text{s}$ resolution as depicted in \cref{fig:psd}.
The Bayes factors for the overtone are shown in \cref{fig:psdbf}.
Using a \ac{PSD} with a less resolved line structure significantly reduces evidence for the overtone. 
Only results using a 2048 Hz sampling rate are presented, but similar conclusions hold across other sampling rates, where the logarithm of the Bayes factors remains below $0$ around the merger time.
A less resolved \ac{PSD} effectively implies a higher noise level, decreasing the likelihood of an overtone model.
This is evident in \cref{fig:psd}, which shows the maximum likelihood waveforms for modes (2,2,0) + (2,2,1).
A $2 \text{s}$ resolution \ac{PSD} tends to favor a lower amplitude for (2,2,1).

\second{We also perform an inference using a \ac{PSD} derived from the Welch method with a 16 s resolution, the results are consistent with that of the 8 s resolution as shown in \cref{fig:psdbf}.}	
We also plot the Bayes factors reported in Refs.~\cite{Cotesta:2022pci,Isi:2022mhy} for comparison.
The Bayes factors derived from the $2 \text{s}$ resolution \ac{PSD} exhibit a higher level of consistency with those from Ref.~\cite{Cotesta:2022pci}. Consequently, we determined that utilizing a shorter data segment, resulting in a less resolved line structure for noise \ac{PSD} estimation, can be one of the contributing factors in diminishing the statistical significance of a ringdown overtone analysis.

\section{Verification by Numerical Relativity Injections}
\label{sec:nr}

\second{
Finally, we apply injections from numerical relativity simulations to validate our \ac{QNM} overtone inference based on gating-and-inpainting.
It is worth mentioning that statistical validation for similar purposes has been conducted in Ref.~\cite{Capano:2022zqm}, using injections from the numerical relativity surrogate model \texttt{NRSur7dq4} \cite{Varma:2019csw}, and in Ref.~\cite{Abedi:2023kot} using \texttt{IMRPhenomTPHM} \cite{Estelles:2021gvs}, a time-domain phenomenological inspiral-merger-ringdown waveform model, to verify the capability of the gating-and-inpainting to effectively recover the \ac{QNM} (3,3,0) mode.
Ref.~\cite{Correia:2023bfn} has also used simulations from a frequency-domain phenomenological template \texttt{IMRPhenomXPHM} \cite{Pratten:2020ceb} to demonstrate the effectiveness when marginalizing over the \ac{QNM} start time.
}

\second{
In this study, we utilize the numerical relativity simulation \texttt{SXS:BBH:0305} from the Simulating eXtreme Spacetimes catalog \cite{Boyle:2019kee, Scheel:2025jct} converted to the convention of LIGO Algorithms Library's coordinate system \cite{Schmidt:2017btt}.
This simulation is designed to resemble the properties of the first detection GW150914.
The detector frame total mass is set to 72 $M_\odot$, with the source located at a luminosity distance of 410 Mpc; the sky location, polarization, inclination, and coalescence phase are the same as the numerical relativity injections in Ref.~\cite{Cotesta:2022pci}.
The component spins, aligned with the orbital angular momentum, are 0.33 and -0.44, respectively.
We inject the simulation to zero noise, hence the inference results should be interpreted as an average over an infinite set of noise realizations, each drawn from the same underlying Gaussian distribution.
We compute the likelihood in a sampling rate of 2048 Hz, still using the same noise \ac{PSD} as in \cref{section:results}.
To further verify our observation about the impact of \ac{PSD} estimation, we also use the \ac{PSD} from a 2s resolution as is discussed in \cref{sec:psd}.
}

\begin{figure}[htbp]
\includegraphics[width=\columnwidth]{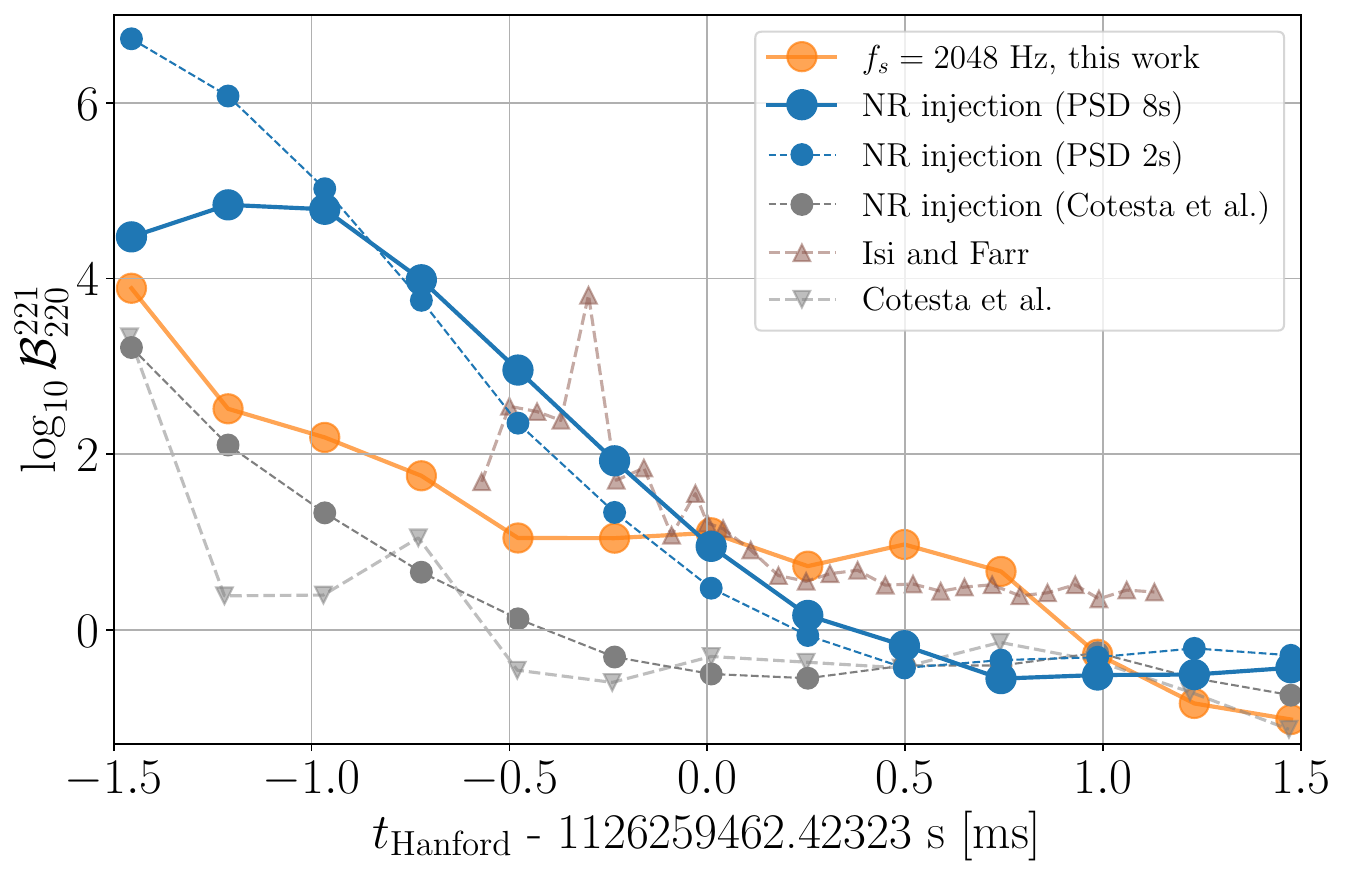}
\caption{The Bayes factors of an overtone drawn from the zero noise injection with a numerical relativity simulation \texttt{SXS:BBH:0305}, using a \ac{PSD} estimation with $8 \text{s}$ and $2 \text{s}$ resolution, respectively.
For comparison, we also plot the results extracted from Ref.~\cite{Cotesta:2022pci} (their Fig. 1) using the same numerical relativity injection into zero noise.
We also replot the Bayes factor reported by this work and that from Refs.~\cite{Cotesta:2022pci, Isi:2019aib, Isi:2022mhy} for a reference.
}
\label{fig:nr}
\end{figure}

\second{
We report our results in \cref{fig:nr}.
As a consequence of zero noise injection, we find that the Bayes factor for an overtone detected in the numerical relativity signal, utilizing a 8s \ac{PSD} resolution, is much smoother than in the real data.
It monotonically decreases as the starting time advances and stabilizes $\sim0.5$ ms after the merger with no evidence of an overtone.
At the merger epoch, the Bayes factor is $\sim9$, consistent with the inferences obtained from the real data.
Just before and after the merger, the Bayes factor decays approximately exponentially, reducing by an order of magnitude when the start time advances by  $\sim 0.25$ ms, again demonstrating the rapid decay of the evidence of an overtone.
For comparison, using a \ac{PSD} with 2s resolution, the Bayes factor is found to be slightly lower around the merger epoch, aligned with our observations in \cref{sec:psd}. 
We also replot the Bayes factor from real data drawn by this work and from Refs.\cite{Cotesta:2022pci, Isi:2019aib, Isi:2022mhy} in \cref{fig:nr} for a reference.
}

\second{
For additional comparison, we extract the results from numerical relativity injection from Ref.~\cite{Cotesta:2022pci} (their Fig. 1), which were derived using the identical SXS simulation injected into zero noise.
These results should be directly comparable to ours obtained with the 2 s \ac{PSD}.
Indeed, we notice a remarkable similarity between the two sets of results, except for an apparent time delay.
If we manually time shift the Bayes factor from \cite{Cotesta:2022pci} by about $0.8$ ms, the Bayes factor results align strikingly well. 
While we can not explain the exact numeric value of the apparent time shift, this provides positive evidence that our analysis using \texttt{PyCBC} inference would converge with the results from \cite{Cotesta:2022pci}, aside from an apparent shift of the time stamps.
}

\section{Conclusion and Discussion}
\label{section:conclusion}
We revisit the GW150914 overtone using an independent \red{gating-and-inpainting} analysis, with settings as similar as possible to \cite{Cotesta:2022pci, Isi:2019aib, Isi:2022mhy}; notably we use a $8~ \text{s}$ data analysis duration with another $4~\text{s}$ both at the head and the rear accounting for the length of the whitening filter.
We examine the frequency spectral content of the recovered waveforms and interpret the divergence of results from different sampling rates as evidence for where the overtone model is not valid to be matched with the data.
When the results converge around the merger time of GW150914, we find Bayes factor values fall within the range of 10 to 26, which supports the existence of an overtone, in agreement with the conclusion of  \cite{Isi:2019aib, Isi:2022mhy}.
We show that starting the analysis too early or too late will lead to discrepancies for different sampling rates.
At too early times, an inappropriately low sampling rate tends to favor a signal with a shorter decay time and thus biases the estimation towards stronger evidence for the $(2,2,1)$ mode.
In light of these discoveries, we propose a new strategy, from the data analysis perspective, by analyzing the convergence of different sampling rates to determine the validity of the overtone model, complementary to the efforts of, e.g., Refs.~\cite{Baibhav:2023clw, Nee:2023osy,Giesler:2024hcr}, which address whether the overtone is valid physically.

We also investigate the systematics of data analysis. The statistical significance of an overtone would be influenced by an imperfect sampling rate, incorrect starting time, and poor resolution of the \ac{PSD} estimation. 
\second{Furthermore, we test the robustness of the Bayes factor results using a numerical relativity simulation.
Additional checks, including changing to another sampling parameter parameterization, changing the gating length or the data analysis duration}, are reported in \cref{app:b}, the results show the Bayes factors remain consistent. 

We summarize our findings below.

(i) Using a low sampling rate, such as 1024 Hz, leads to inconsistent overtone significance when starting the analysis before the merger, as demonstrated in \cref{fig:bf}.
Therefore, we propose a \second{consistency test from different sampling rates} to identify the valid region for an overtone.
A sampling rate of $2048$ Hz or higher is sufficient for overtone analysis when starting at or after the merger.

\red{(ii) Given the extremely short time of an overtone (e.g., with a final mass of 60 $M_\odot$ and a final spin of 0.8 leading to a decay time of 1.3 ms for the (2,2,1) overtone), rounding to the nearest discretized data sample yields an inconsistent and inaccurate starting time across two or more \ac{GW} detectors. This, in turn, biases the statistical evidence, as illustrated in \cref{fig:timeshift} with a 2048 Hz sampling rate example. Hence, we accurately determine the starting time by reconstructing the subsample.}

\red{(iii) Using a short data segment to estimate the \ac{PSD} of noise does not resolve the line structure well in the frequency domain, effectively increasing the noise level and reducing the evidence for the overtone model, as demonstrated in \cref{fig:psdbf}. Therefore, we adopt an $8~\text{s}$ long data segment to resolve the \ac{PSD} line structure for overtone analysis.}

Recently, the authors of Refs.~\cite{Isi:2019aib, Isi:2022mhy} published a comment showing that increasing the analysis duration and correcting the approximation of starting time discretization can alleviate the discrepancies \cite{Isi:githubnote}.
However, the authors of \cite{Cotesta:2022pci} replied that the logarithmic Bayes factors are still negative after addressing the comments  \cite{reply_Carullo_etal}.
Our method in the current work does not have any of the aforementioned limitations as we have used a long analysis duration ($16$ s) and reconstructed the subsampling data point to ensure the starting time is precise (\cref{app:a}).
Our work shows that using these more robust choices will affect the results in the direction of enhancing the statistical significance of an overtone.

We release the scripts to reproduce this work and the posterior files at \cite{github}.

\acknowledgments
Y.-F.W. acknowledges the Max Planck Gesellschaft, the Atlas computing team, and AEI Hannover where the majority of this work was done; he also thanks Gregorio Carullo, Max Isi and Will Farr for insightful comments. J.A. was supported by ROMFORSK grant Project No. 302640.
C.D.C. acknowledges support from NSF award PHY-2309356.
A.H.N. acknowledges support from NSF grant PHY-2309240. 
S.K.\ acknowledges support from the Villum Investigator program supported by the VILLUM Foundation (grant no.\ VIL37766) and the DNRF Chair program (grant no.\ DNRF162) by the Danish National Research Foundation. This project has received funding from the European Union's Horizon 2020 research and innovation programme under the Marie Sklodowska-Curie grant agreement No 101131233.

This research has made use of data or software obtained from the Gravitational Wave Open Science Center (gwosc.org), a service of the LIGO Scientific Collaboration, the Virgo Collaboration, and KAGRA. This material is based upon work supported by NSF's LIGO Laboratory which is a major facility fully funded by the National Science Foundation, as well as the Science and Technology Facilities Council (STFC) of the United Kingdom, the Max-Planck-Society (MPS), and the State of Niedersachsen/Germany for support of the construction of Advanced LIGO and construction and operation of the GEO600 detector. Additional support for Advanced LIGO was provided by the Australian Research Council. Virgo is funded, through the European Gravitational Observatory (EGO), by the French Centre National de Recherche Scientifique (CNRS), the Italian Istituto Nazionale di Fisica Nucleare (INFN) and the Dutch Nikhef, with contributions by institutions from Belgium, Germany, Greece, Hungary, Ireland, Japan, Monaco, Poland, Portugal, Spain. KAGRA is supported by Ministry of Education, Culture, Sports, Science and Technology (MEXT), Japan Society for the Promotion of Science (JSPS) in Japan; National Research Foundation (NRF) and Ministry of Science and ICT (MSIT) in Korea; Academia Sinica (AS) and National Science and Technology Council (NSTC) in Taiwan.
\vspace{5mm}

\appendix

\section{Proof for the solution of inpainting}
\label{sec:proof}
Recall that the Gaussian likelihood for the noise \boldn{} is
\be
\ln \mathcal{L} = -\frac{1}{2} \bold{n}^TC^{-1}\bold{n} \red{ -\ln\sqrt{(2\pi)^N|C|}}.
\ee
As firstly introduced in \cite{Zackay:2019kkv}, they construct an inpainting operator 
\be
\label{eq:inpaintingfilter}
F = \bold{1} - AM^{-1}A^T C^{-1}
\ee
where $M = A^T C^{-1} A$. 
The matrix $A$ is an ``extraction matrix" with the size $N \times M$, where $N$ and $M$ are the numbers of elements of \boldn{}, and $\bold{x}$, the bad data to be inpainted, respectively.
Explicitly, A is an identity matrix in the rows corresponding to $\bold{x}$ and zeros elsewhere
\be
A= \begin{pmatrix}
~ & ~ &  0  & ~  & ~ \\
~ & 1 & ... & ~  & ~ \\
~ & ~ &  1  & ...& ~ \\
~ & ~ &  ~  & 1  & ...  \\
~ & ~ & ... & ~  & ~ \\
~ & ~ &  ~  & ... & 1 \\
~ & ~ &  0  & ~  & ~
\end{pmatrix}
\ee
Such construction will have the desirable property that, after acting $F$ on \boldn{},  any elements in the gating region will not impact the computation of $\bold{n}^T C^{-1} F\bold{n}$. 

We offer another perspective, which is mathematically equivalent to \cite{Zackay:2019kkv}, by considering the inverse of the covariance from the truncated data.
Without loss of generality, we express $\bold{n}_\mathrm{inpaint}$ as the concatenation of the truncated data and the (yet unknown) inpainting data, $\bold{n}_\mathrm{inpaint} = \bold{n}_\mathrm{tr} \oplus \bold{x}$ (the more general case that $\bold{x}$ is in the middle of $\bold{n}_\mathrm{inpaint}$ can be obtained by acting permutation matrix on it, and the following derivation remains the same).
Hence the covariance matrix can be formally expressed by a block matrix, its inversion is
\be
C^{-1} = 
\begin{pmatrix}
C_\mathrm{tr} & B \\
B^T & D
\end{pmatrix}^{-1} = 
\begin{pmatrix}
a & b \\
b^T & d
\end{pmatrix}
\ee
where $B, D, a, b, d$ are all block matrices yet unknown. 
Because of the inversion relation, we have
\begin{eqnarray}
\label{eq:matrixinv}
C_\mathrm{tr} a + Bb^T = \bold{1} \\ \nonumber
C_\mathrm{tr} b + Bd = \bold{0}
\end{eqnarray}
where $\bold{1}$ and $\bold{0}$ are the unity and zero matrix, respectively. 
Express $B$ by the second line of Eq. (\ref{eq:matrixinv}) and insert to the first line, one gets 
\be
C_\mathrm{tr}^{-1} = a-bd^{-1}b^T
\ee
Hence the likelihood of truncated data can be written as
\be
\label{eq:proof}
\bold{n}_\mathrm{tr}^T C_\mathrm{tr}^{-1} \bold{n}_\mathrm{tr} = 
\bold{n}_\mathrm{tr}^T a \bold{n}_\mathrm{tr} - \bold{n}_\mathrm{tr}^Tbd^{-1}b^T \bold{n}_\mathrm{tr}
\ee
In the main text we have introduced the solution to be
\be
\label{eq:appendixtoe}
C^{-1}(\bold{n}_\mathrm{tr} \oplus \bold{x})_\mathrm{inpaint} = \bold{0}_\mathrm{inpaint}
\ee
This can be expressed as
\be
b^T \bold{n}_\mathrm{tr} + d \bold{x} = \bold{0}
\ee
Hence
\be
\bold{x} = - d^{-1} b^T \bold{n}_\mathrm{tr}
\ee
Therefore
\begin{eqnarray}
&& (\bold{n}_\mathrm{tr} \oplus \bold{x})^T C^{-1}(\bold{n}_\mathrm{tr} \oplus \bold{x})  \\ 
 &=& (\bold{n}_\mathrm{tr} \oplus \bold{x})^T \begin{pmatrix}
a & b \\ \nonumber
b^T & d
\end{pmatrix}(\bold{n}_\mathrm{tr} \oplus \bold{x}) \\  \nonumber
 &=& \bold{n}_\mathrm{tr}^T a \bold{n}_\mathrm{tr} - \bold{n}_\mathrm{tr}^T bd^{-1}b^T \bold{n}_\mathrm{tr}  
\end{eqnarray}
Together with \cref{eq:proof} we have proved that inpainting with $\bold{x}$ will resume the use of $C^{-1}$ in the likelihood. 
As discussed in \cite{Isi:2021iql}, constructing the inpainting filter directly as in \cref{eq:inpaintingfilter} invokes inverting the $A^TC^{-1}A$ which requires $M^3$ time complexity. 
However, as we use \second{Levinson-Durbin Recursion} \cite{levinson1946wiener,durbin1960fitting} in \texttt{scipy} \cite{Virtanen:2019joe} to solve the Toeplitz linear equation for inpainting data in \cref{eq:appendixtoe}, it only requires the $M^2$ time complexity.

Moreover, we verified the robustness of our gating-and-inpainting solver by confirming that the over-whitened waveform, $h(f)/S_h(f)$, is indeed zero within the inpainted area, as expected by solving \cref{eq:toeplitz}.

\section{Checks on the impacts of ringdown amplitude parameterization, data analysis duration and gating length}
\label{app:b}

To verify the robustness of our inference configurations, we performed various checks by sampling over another overtone amplitude parameterization, adjusting the data analysis duration, and the length of gating-and-inpainting.
The results for the evidence of an overtone are consistent with the baseline models used to present our main results in \cref{section:results}.

\second{
We first consider another overtone amplitude parameterization.
When sampling the likelihood distribution, we notice that the label switching issue between the $(2,2,0)$ and $(2,2,1)$ mode would sometimes occur, in which the sampler would explore where the $A_{220}$ is almost zero, and $A_{221}$ is favored associated with much heavier remnant mass.
This is because the $(2,2,1)$ mode of the template is locked on the $(2,2,0)$ signal in the data.
}

\second{
In light of this issue, we choose to sample the relative amplitude between the (2,2,1) and (2,2,0) mode, $A_{221}^\mathrm{rel} = A_{221}/A_{220}$, instead of the absolute amplitude.
}
We choose $A_{221}^\mathrm{rel}$ to be uniform in $[0, 5]$, and $\log_{10} A_{220}$ uniformly distributed in $[-24,-19]$.

\begin{figure}[htbp]
\includegraphics[width=\columnwidth]{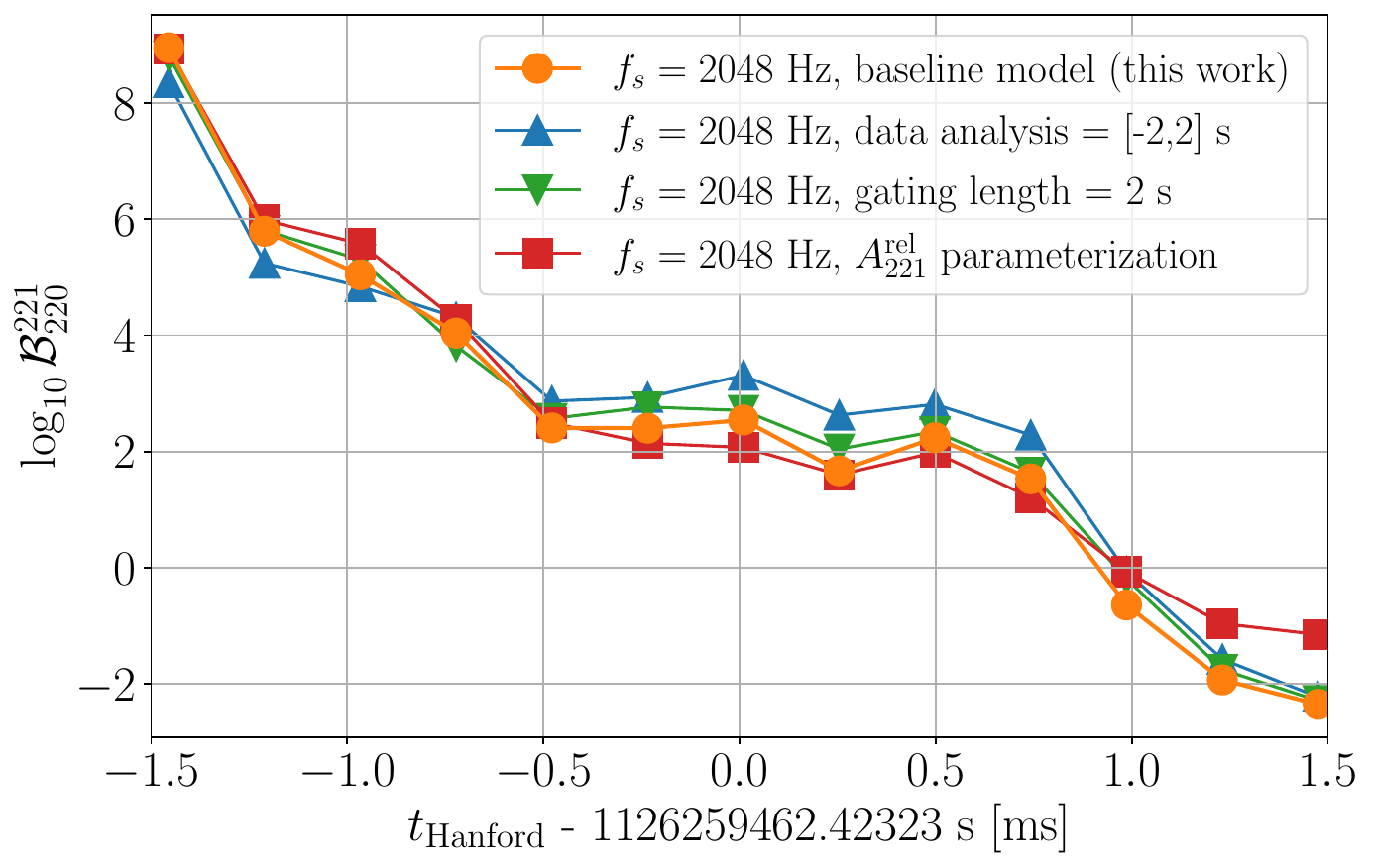}
\caption{Comparison of Bayes factors for a different parameterization that samples $A_{221}^\mathrm{rel}$ and $\log_{10} A_{220}$, a different data analysis duration and a different gating-and-inpainting length.
The results with adjusted settings are consistent the baseline settings used to report the main results of this work.}
\label{fig:diffpar}
\end{figure}

The results of Bayes factors are shown in \cref{fig:diffpar}.
We only show the results from $f_s = 2048$ Hz; other sampling rates present consistent conclusions. 
We note that the two parameterizations agree with each other well.
At zero epoch, the relative amplitude parameterization slightly prefers a lower $\mathcal{B}^{221}_{220}$, which can be attributed to a slightly higher weight in the prior space for $A_{221}=0$.
At a sufficiently late time, the relative parameterization favors a Bayes factor being 1 between the $(2,2,1)+(2,2,0)$ and $(2,2,0)$ modes, i.e., no preference for any one of the models.

\second{
We have also changed the data analysis duration from $[-4, 4]$ s around the merger epoch of GW150914 used in \cref{section:results} to $[-2, 2]$ s, or changed the gating-and-inpainting length from $1$ s to 2 s.
The Bayes factors for an overtone are presented in \cref{fig:diffpar} and broadly consistent with the baseline settings.
Overall, the results demonstrate the robustness of the Bayes factor results in the main text.
}

\bibliography{ringdown.bib}
\end{document}